\documentclass{emulateapj}
\usepackage{epsfig}
\usepackage{natbib}

\def\apj{ApJ}
\def\aap{A\&A}
\def\apjl{ApJL}
\def\apjs{ApJS}
\def\mnras{MNRAS}

\def\araa{ARA\&A}
\def\nat{Nature}

\newcommand{\tmax}{T_{\rm max}}

\newcommand{\K}{\,{\rm K}} 
\newcommand{\msun}{M_{\odot}} 
\newcommand{\hinv}{h^{-1}}

\newcommand{\lcdm}{$\Lambda$CDM }

\newcommand{\hkpc}{h^{-1}\;{\rm kpc}}

\begin{document}
\title{Seeding the formation of cold gaseous clouds in Milky Way size halos}
\author{Du\v{s}an Kere\v{s} \& Lars Hernquist}
\affil{Harvard-Smithsonian Center for Astrophysics, Cambridge, MA 02138; \texttt{dkeres@cfa.harvard.edu}}

\begin{abstract}

We use one of the highest resolution cosmological SPH simulations to
date to demonstrate that cold gaseous clouds form around Milky Way
size galaxies.  We further explore mechanisms responsible for their formation
and show that a large fraction of clouds originate as a consequence of
late-time filamentary ``cold mode'' accretion.  Here, filaments that are 
still colder and denser than the surrounding halo gas,   
are not able to connect directly to galaxies, as they do at high
redshift, but are instead susceptible to the combined action of
cooling and Rayleigh-Taylor instabilities at intermediate radii within
the halo leading to the production of cold, dense pressure-confined
clouds, without an associated dark matter component. This process is
aided through the compression of the incoming filaments by the hot halo
gas and expanding shocks during the halo buildup.
Our mechanism directly seeds clouds from gas with substantial
local overdensity, unlike in previous models, and provides a channel for the 
origin of cloud complexes.  These 
clouds can later ``rain'' onto galaxies, delivering fuel for star
formation.  Owing to the relatively large cross section of filaments
and the net angular momentum carried by the gas, the clouds will be
distributed in a modestly flattened region around a galaxy.

\end{abstract}


\section{Introduction}

Observations of neutral hydrogen around the Milky Way (MW) have
revealed that our Galaxy is surrounded by a large number of cold
gaseous clouds.  Owing to their high velocities compared to galactic
rotation, it was proposed that these objects reside outside the disk
\citep{muller63}, and they became known as high-velocity clouds
(HVCs).  The masses of most HVCs are uncertain because of their 
uncertain distances (several HVC complexes are now placed at $D\sim10kpc$ 
\citep{wakker08, thom08}), but are estimated to be $10^5-10^7\msun$
\citep{wakker97}.  Similar HI clouds and cloud complexes have been
found around other galaxies \citep[for a review see][]{sancisi08}.

The overall kinematics of HVCs implies net inflow, motivating a
picture where HVCs represent the building blocks of our Galaxy
\citep{oort70} and provide the supply of gas needed for its relatively
high star formation rate.  In fact, our Galaxy has been forming stars
at a steady rate for the last several Gyrs \citep{binney00}, implying
that the gas consumed must be continuously replenished.

Many HVCs can be ionized by the external UV field, so HI observations
likely detect only the densest of these structures that are able to
self-shield.  The presence of clouds around other galaxies is also
implied by absorption line studies at both low and intermediate
redshifts \citep{lanzetta90, gauthier09}, which typically find gaseous
clouds to distances of several tens of kiloparsecs in a halo.  It is
likely that these absorbing clouds are analogs to the local HVC
population \citep{maller04}.

Despite numerous theories,
the origin of these clouds remains uncertain.  Proposed mechanisms
fall into the following three classes: 1: Cooling instabilities in the
hot medium \citep{field65, mo96} that can interact with the hot halo
gas resulting in net infall onto the Milky Way \citep{maller04}.  2:
Production by a ``galactic fountain'' where outflows driven by
supernovae from the disk condense in the hot corona and fall back onto
the MW \citep{bregman80}, possibly entraining hot gas from the halo
\citep{fraternali08}.  3: Interaction of hot halo gas with infalling
satellites \citep[e.g.][]{putman03b}, to account for clouds with
specific properties; e.g. those in the Magellanic Cloud system.  In
principle, all of these channels could operate, complicating modeling
and data interpretation. Here, we concentrate on cloud formation
  from cooling and hydrodynamical instabilities in infalling gas.    

Cosmological hydrodynamic simulations have shown that at high redshift
the bulk of the gas supplied to galaxies is accreted from the
intergalactic medium through filaments in a ``cold mode''
\citep{katz03, keres05, mythesis, keres09a, dekel06, ocvirk08,
brooks09} even after a hot halo atmosphere develops. Owing to lower
densities, at late times there is a shift in the physics that governs
the accretion, where filamentary flows get disrupted in the central
parts of massive halos.  We examine this late-time accretion and show
that these disrupted filaments are still able to supply cold gas to MW
size galaxies.  We demonstrate that accretion through filaments
creates density inversions in halos, triggering a combination of
cooling and Rayleigh-Taylor (RT hereafter) instabilities that produce
clouds, supplying MW like galaxies with gas.  Earlier work hinted that
the ``leftovers'' from filamentary might contribute to cloud
formation \citep{sommer-larsen06, connors06, peek08, keres09a} but did
not demonstrate why and how this process operates.

\section{Simulations}

We use the Gadget-2 Smoothed Particle Hydrodynamics (SPH) code
\citep{springel05a} which simultaneously conserves energy and entropy,
preventing numerical phase mixing.  This is potentially an important
property for modeling cloud formation as many SPH implementations can
mix cold, dense and hot, dilute phases \citep{springel02,
mythesis}. 
Gravitational forces are calculated combining Particle Mesh
and hierarchical tree algorithms.  In order to capture the formation
of $\sim10^6\msun$ clouds at high spatial resolution in a large
volume, we adopt a zoom-in technique, where we first simulate a
cosmological box of $10\hinv$ Mpc on a side using only dark matter
(DM hereafter) particles. Next, we select a region that will collapse
into a MW size object by z=0. We populate this region and its
surroundings at high resolution with gas and DM particles, introducing
\lcdm density fluctuations down to the new inter-particle
  separation scale. Outside of this region, the density field is
sampled more coarsely with high mass DM particles to provide long
range forces. Such initial conditions are re-generated at $z=99$ using
the \citet{eisenstein99} power spectrum in a \lcdm cosmology: $\Omega_b=0.044$,
  $\Omega_m=0.27$, $h=0.7$, $\sigma_8=0.8$.
We include standard cooling processes for a gas of
primordial composition \citep{katz96} and the UV background
\citep{haardt96}.  For simplicity, we neglect galactic outflows.

In total, the simulation has 6.6 million particles. The most massive
halo in the high-resolution region (hereafter, MWH), analyzed in
detail in what follows, at z=0 has $M_{halo}=7\times10^{11} \msun$
and contains 1.4 million baryonic and 0.9 million DM particles.  Each
gas particle begins with a mass of $9 \times 10^4h^{-1} \msun$, while
stars are half this mass.  Gravitational softening is the same for DM
and baryonic particles, $\epsilon =400\hinv (1+z)^{-1}$pc
(Plummer equivalent).

\section{Results}
\subsection{Clouds}

\begin{figure*}
\epsfxsize=3.4in
\epsfbox{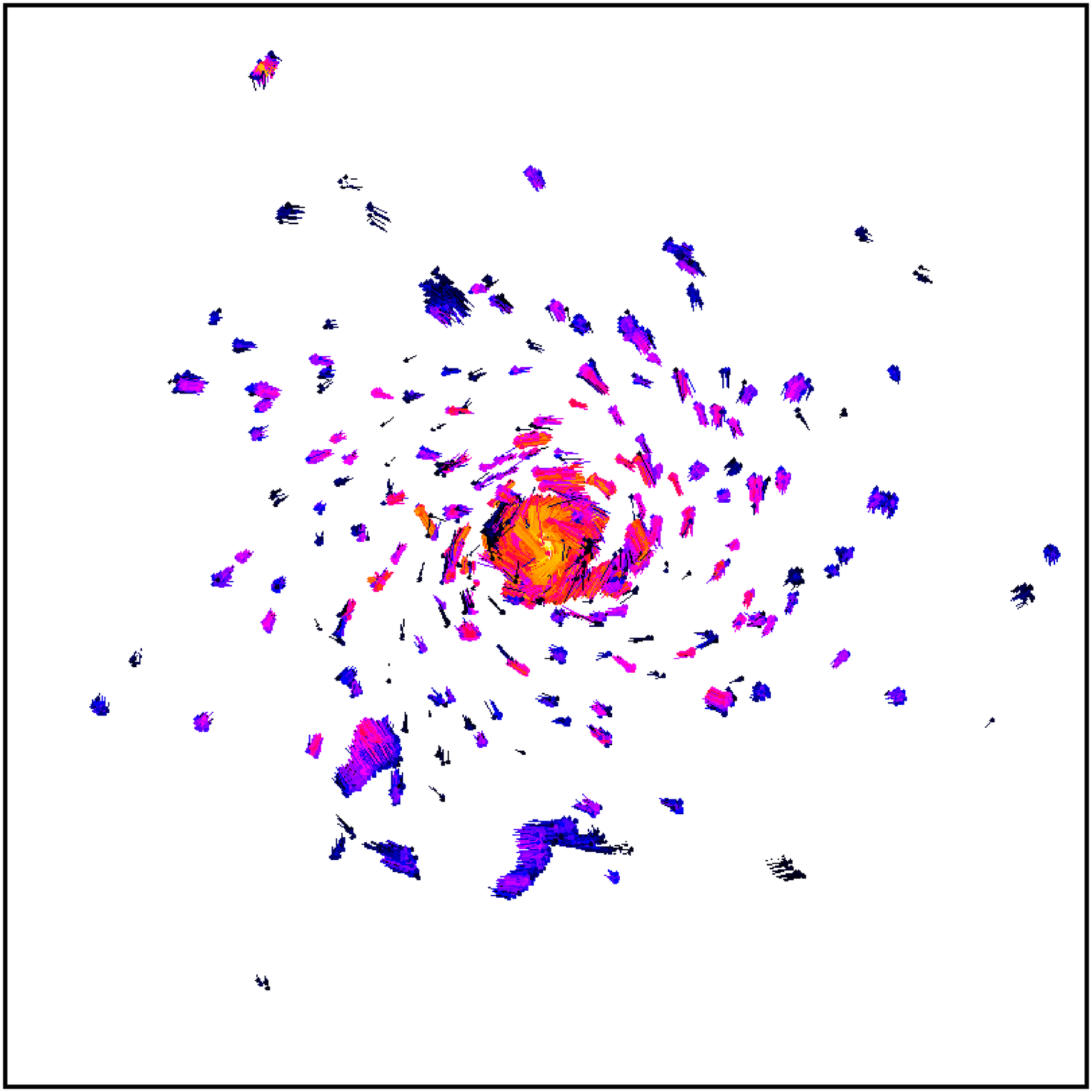}
\epsfxsize=3.4in
\epsfbox{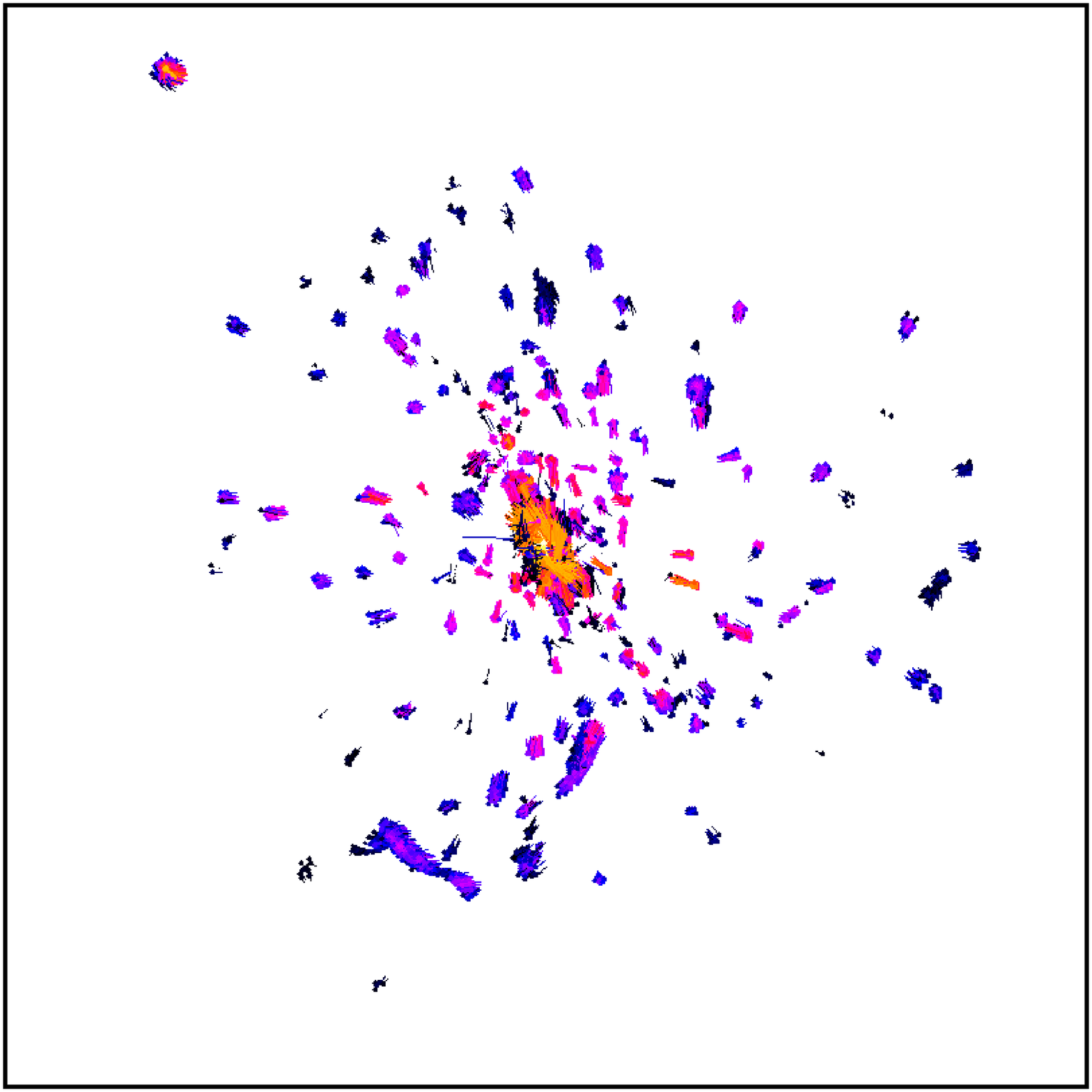}
\caption{Distribution of dense clouds in a 
  $200\hinv$ kpc box centered on the most massive galaxy in MWH. 
  Particles are color
  coded by density from $n_H=4\times10^{-4}$ cm$^{-3}$ (darkest blue) to $4$
  cm$^{-3}$ (bright yellow). Clouds have temperatures $T_c\sim10^4 \K$.
}
\label{fig:clouds}
\end{figure*}

In Figure~\ref{fig:clouds}, we show the inner regions of MWH.  Gas
particles with $n_H > 4\times10^{-4} cm^{-3}$ are plotted close to
face-on and edge-on projections of the disk, revealing the presence of
a population of dense gaseous clouds together with a central galactic
disk.  Plotting gas colder than 30000K selects an identical cloud
population.  At low redshift MWH contains a large
number of cold, $T_c\sim10^4K$, dense clouds in a $40\hkpc$ (edge
on) times $60\hkpc$ (face on) region around the disk.  We plot gas particles 
with small arrows to indicate the velocity field, demonstrating that the
clouds exhibit a global sense of rotation, similar in orientation to
the outer disk.

We identify clouds with $T < 30000 \K$ and $n_H > 0.004$ cm$^{-3}$
using the code SKID. The results are mainly determined by the
temperature threshold and are robust for any $T < 10^5 \K$.  To
distinguish clouds from galaxies we modified SKID to include thermal
support of gas to properly identify self-gravitating systems.  Bound
objects include the main galactic disk, several satellite galaxies
(one visible in the upper left corner of Figure~\ref{fig:clouds}), and
a small number of gaseous clumps that are not associated with a DM or
stellar component. These self-gravitating clouds are close to the halo
center, where the external pressure compresses them to high densities.
For simplicity, we exclude all bound objects from the remainder of the
analysis, but apart from the main gaseous disk they contribute
negligibly to the cold gas budget.  Cloud masses range from $10^6$ to
few $10^7\msun$. We note, however, that we are limited at low masses
by the simulation resolution.  Some of the most massive clouds are at
larger radii and often represent a dense slab of gas, which might be
susceptible to further instabilities.  Typical column densities (for
the total cold gas, i.e. no ionization correction which could be
substantial) of clouds are $N_H\sim5\times10^{18} - 10^{21}
cm^{-2}$.

\subsection{Gas in the inner halo}

The amount of virialized gas increases with time as the halo grows and
hot gas fills low density regions in-between colder, denser filaments
\citep{keres05}.  At high redshift, filaments survive within hot halos
even when the hot medium dominates the halo gas budget ($M_h > 3\times
10^{11}\msun$).  However, at lower redshift the density of filaments
decreases and the temperature increases so they are more easily
destroyed at small radii by large scale shocks in a hot halo.  This
happens where halo shock conditions are favorable for hot, overdense
gas to propagate outwards \citep{birnboim03}.  Filamentary remnants
that penetrate into halos but not quite to galaxies set the stage for
cloud formation.  For the MWH this epoch starts roughly at $z=1-1.5$
after the last major merger is completed. The last significant ($\sim1:7$) 
merger occurs around z=0.5 and soon afterwards MWH reaches a quasi-static
configuration that lasts 2-3 Gyrs, although cloud formation proceeds
in a similar manner for $z < 1$. At $z\sim0$ around 10 percent of the
halo gas (excluding gas in the galaxies) within the inner $100 \hkpc$,
$1.5-2\times 10^9 \msun$, is in cold gaseous clouds as a result of a
balance between cloud creation, accretion and destruction. This is
consistent with estimates of the total mass of the HVC population
\citep{putman06}. The density of the hot component at a distance of 50
kpc from the main galaxy is $n_H\sim0.5-1.5 \times 10^{-4}cm^{-3}$,
consistent with observational limits \citep{grcevich09}.

\subsection{The origin of the cold clouds}

\begin{figure}
\centerline{
\epsfxsize=0.9\columnwidth
\epsfbox{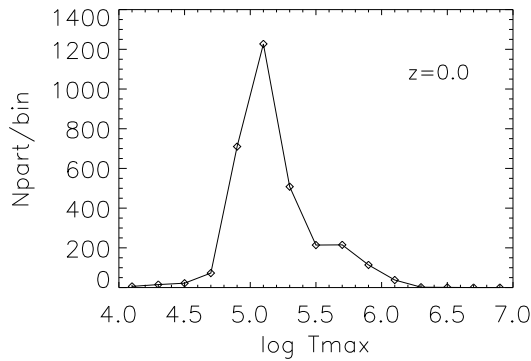}
}
\caption{Histogram of the maximum temperature cloud particles reached
  before condensing into clouds, in bins of $0.2 \log{\tmax}$.
}
\label{fig:tmax}
\end{figure}

To understand their origin we follow cloud particles backwards in time
and determine the maximum temperature that they had in the past,
$\tmax$. This tells us if the clouds form through instabilities in a
hot corona or if they were seeded by colder structures. Figure
~\ref{fig:tmax} shows the results of this analysis (for ``resolved''
clouds, $\geq $ 32 particles).  Most of the mass in the clouds never
reached $0.5-1.5\times 10^6 K$, characteristic of the MWH at z=0, but
instead clouds form from, $1-1.5\times 10^5 K$ material, before they
cool to $\sim10^4K$.  These are temperatures characteristic of
low-redshift, large-scale filaments fueling gas onto MWH.  The
distribution of $\tmax$ is similar in the redshift interval $0 < z
<1$, implying a connection between cloud formation and ``cold mode''
accretion active at higher redshifts.

\begin{figure*}
\begin{center}
\epsfxsize=3.4in
\epsfbox{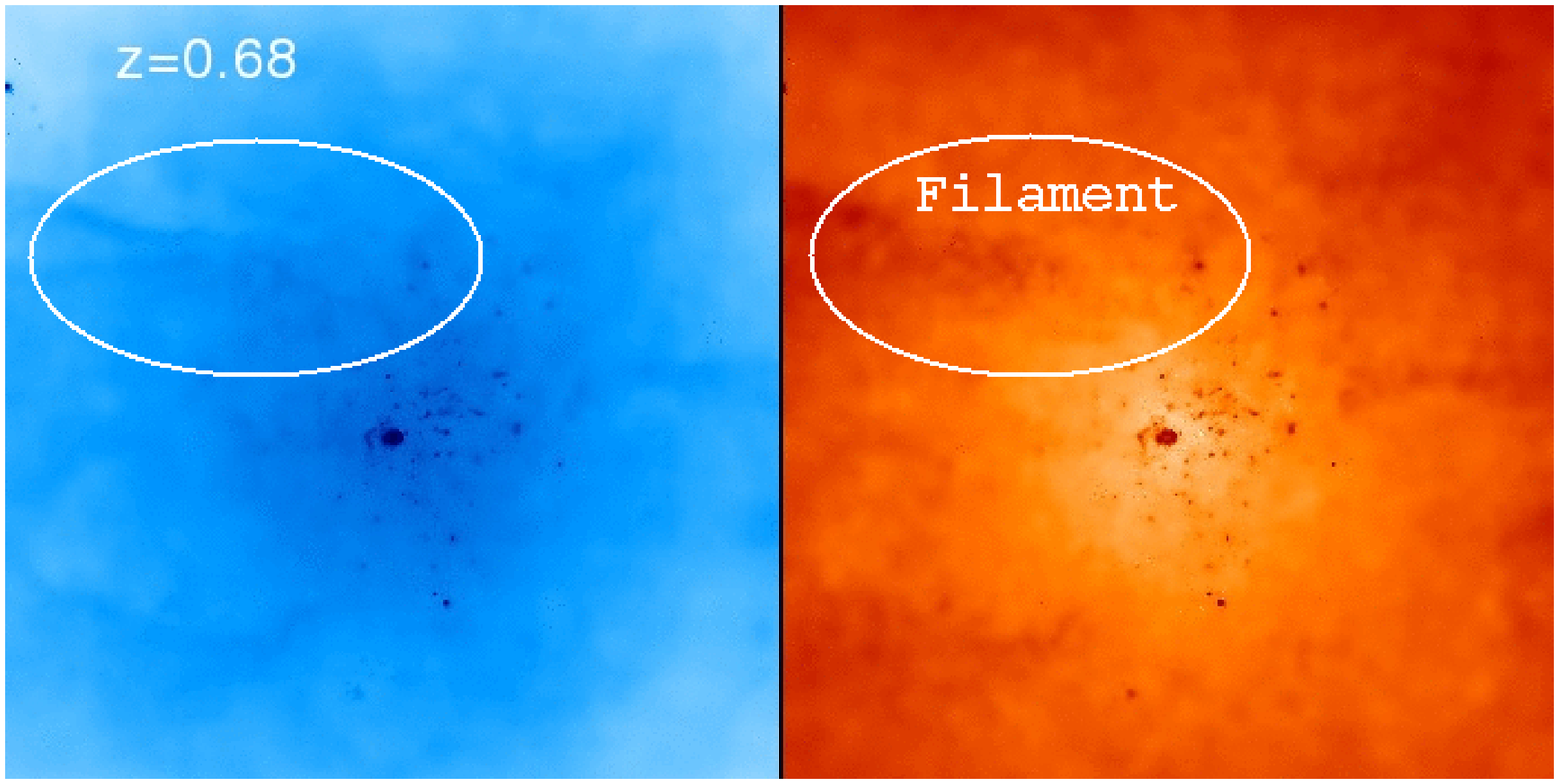}
\epsfxsize=3.4in
\epsfbox{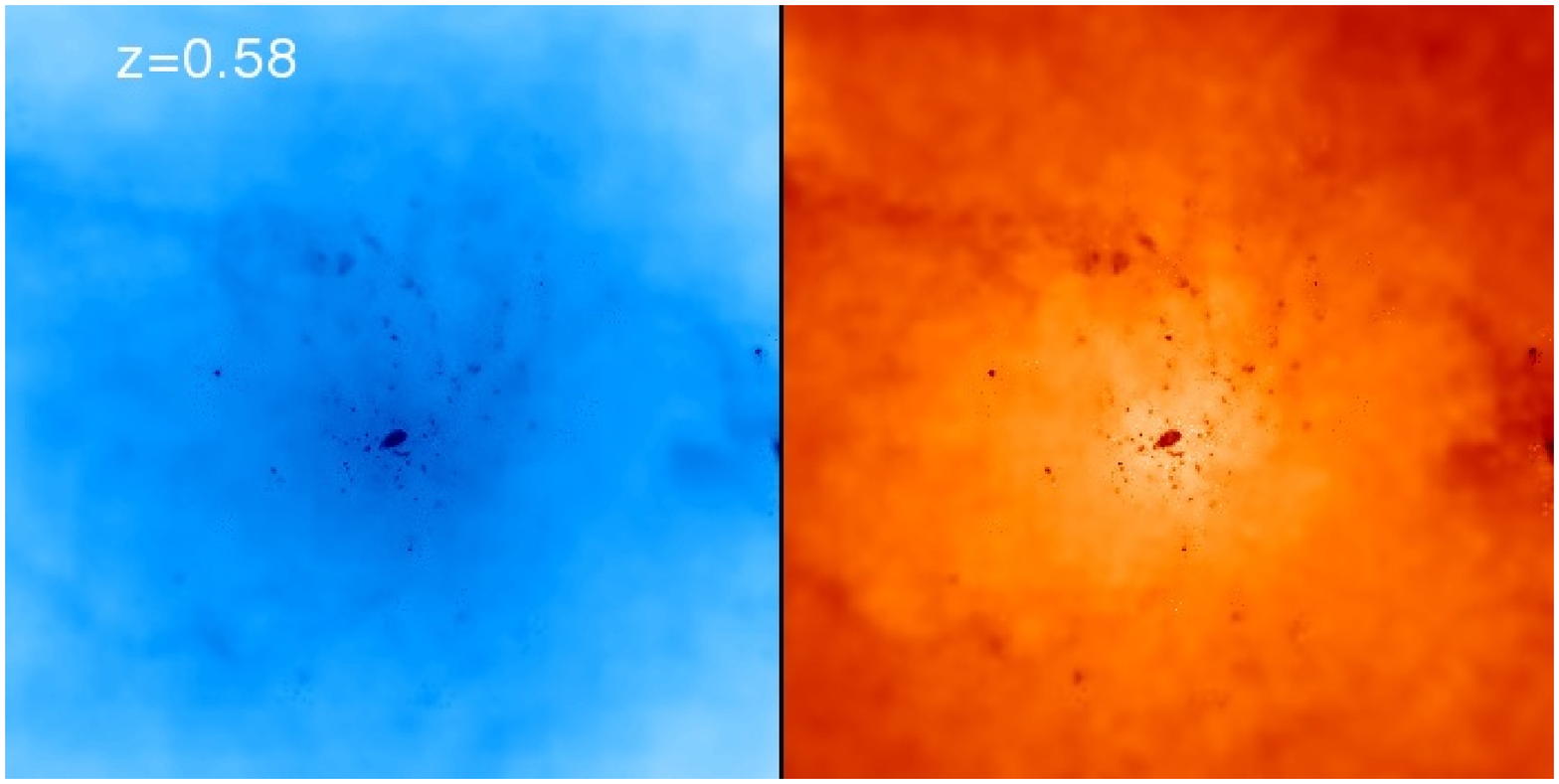}
\epsfxsize=3.4in
\epsfbox{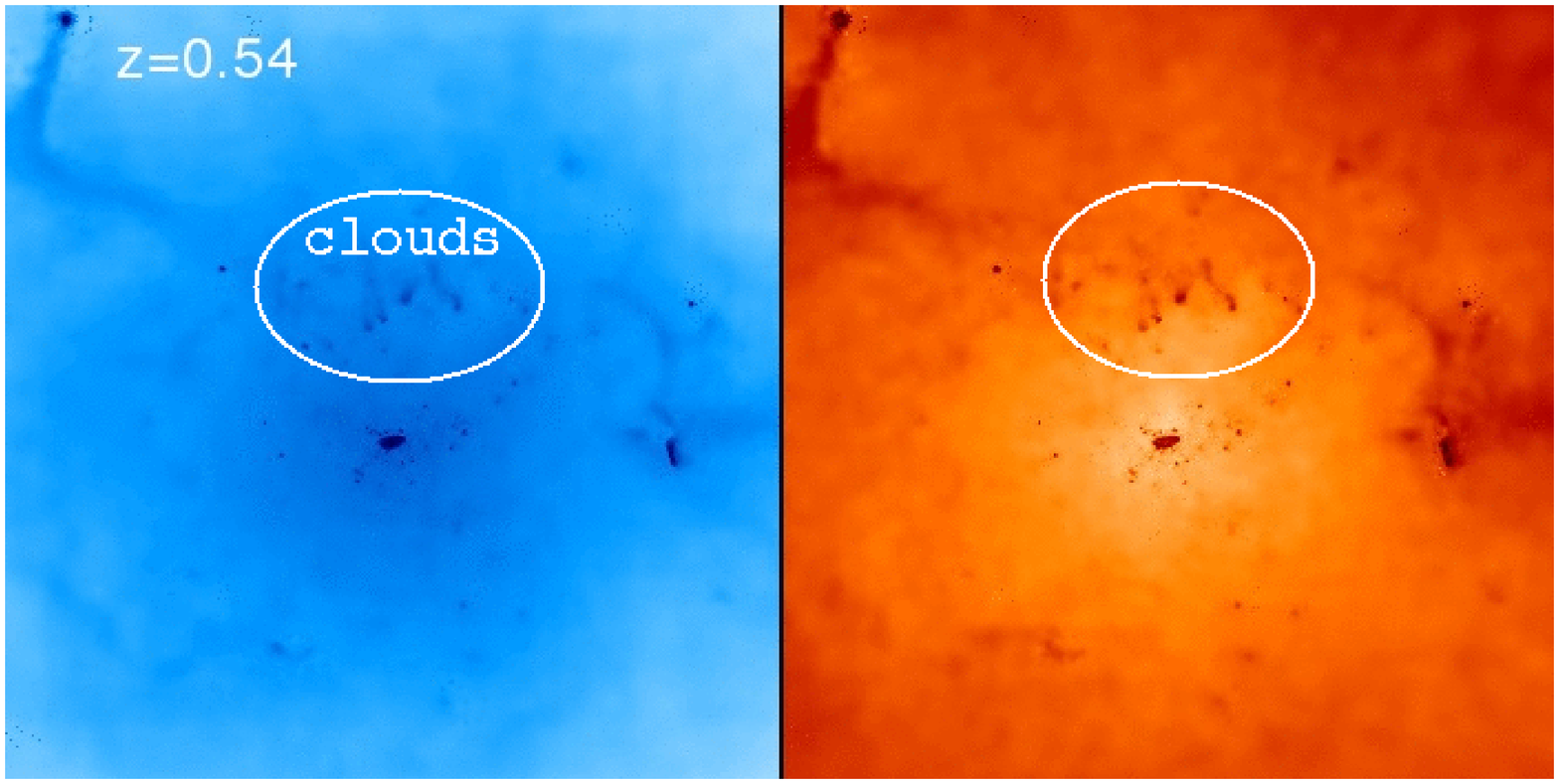}
\epsfxsize=3.4in
\epsfbox{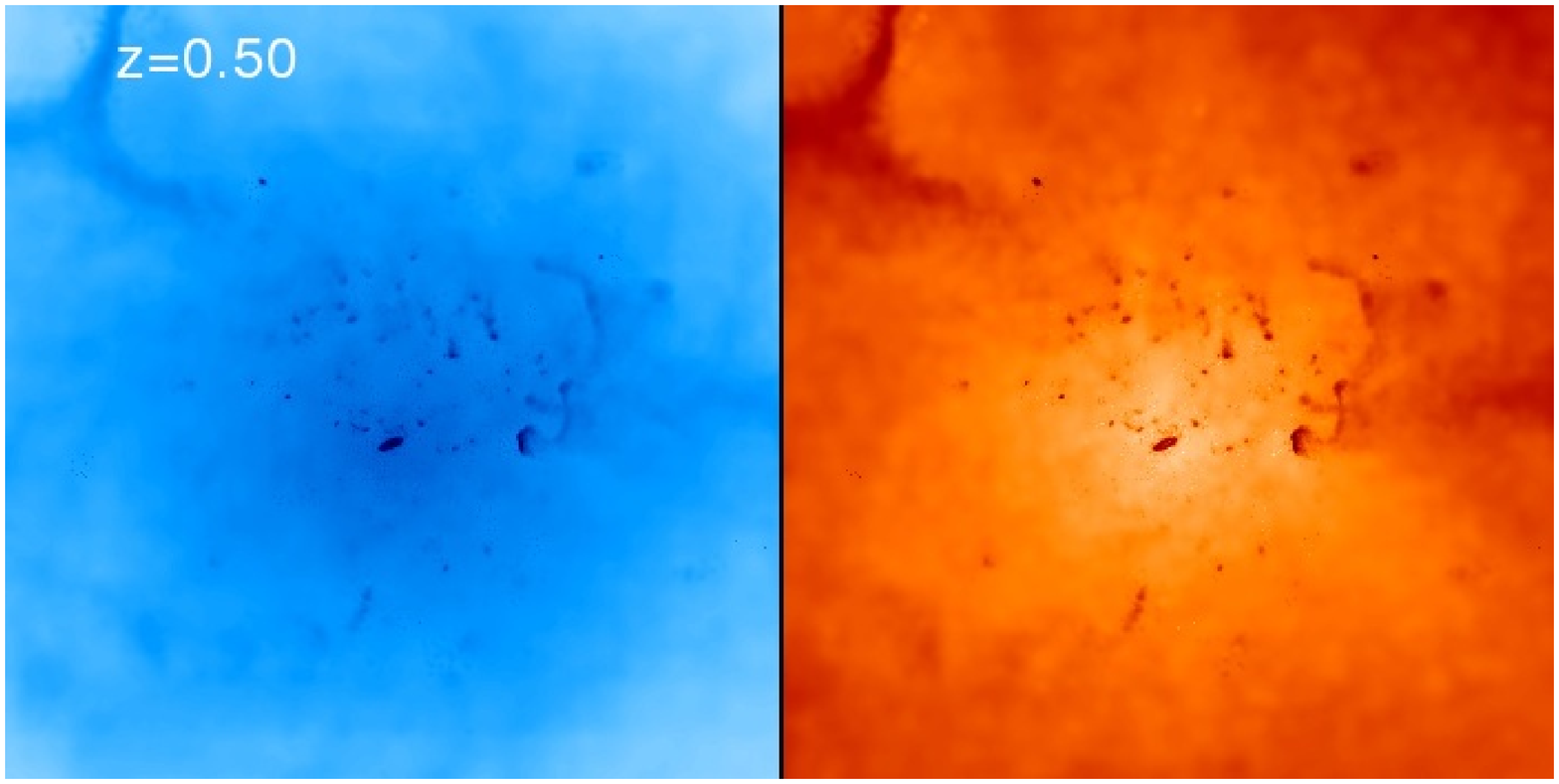}
\end{center}
\caption{Example of the instability developing from an overdense
  filament in a hot dilute halo gas leading to cloud formation. Left
  panels show the projected density and the right panels show projected
  mass-weighted temperature in a box $200 \hkpc$ on a side and
  $100 \hkpc$ thick.}
\label{fig:RT}
\end{figure*}

We illustrate this process in Figure~\ref{fig:RT}.  An overdense, warm
filament penetrates to intermediate radii in the halo, visible in the
panels which show density and temperature. The filament is not
directly connected to the galaxy but manages to reach 30-40 kpc radius
before being heated to higher temperatures.  The large impact
parameter with respect to the central galaxy creates a density
inversion where ``heavier'' fluid lies on top of a ``lighter'' one in
an external gravitational field (of the parent halo), providing the
conditions for RT instabilities to operate.  The initial instabilities
that act on overdense filamentary gas often produce cometary tail-like
morphologies, similar to RT instabilities in the ISM
\citep[e.g.][]{odell97}.  These are further compressed by higher
pressure gas at slightly smaller radii and quickly become cooling
unstable, forming cold, dense clouds.  Additional sites of cloud
formation throughout the MWH can be seen in the figure, often where
compression of infalling gas by halo shocks and pressure from 
surrounding hot gas promotes faster development of instabilities.

The filament is also moving with respect to the surrounding lower
density gas. It is therefore possible that Kelvin Helmholtz (KH)
instabilities can affect this configuration.  It has been shown that
SPH techniques have difficulty handling KH instabilities
\citep{agertz07} so it is important to estimate the timescales for a
dominant mode to affect and destroy such penetrating filaments before
they become cooling or RT unstable at small radii. The relevant
timescale is:
\begin{equation}\label{eq:KH}
t_{KH}= {\lambda (\rho_c+\rho_h) \over {(\rho_c\rho_h)}^{1/2}U} \sim
\lambda {(\rho_c/\rho_h)}^{1/2}/U \, ,
\end{equation}
where $\rho_c$ and $\rho_h$ are the densities of the heavier and more
dilute components, respectively, $\lambda$ is the spatial scale
of the instability, and $U$ is the relative velocity of the two phases.  
For a spherical cloud moving in a dilute medium the
disruption time is $2-5\times t_{KH}$ (self-gravity can make it even
longer) \citep{murray93}.
At the temperatures typical of penetrating filaments in the $z\sim0.5$ MWH, 
$T\sim10^5 K$ which for approximate pressure equilibrium implies
a density contrast of $\sim5-10$. 
The overdense region in Figure~\ref{fig:RT} moves with velocities $\sim100$
km/s through the hot medium. 
The largest available mode is responsible for the destruction of structures.
The length of the structure is $\sim100$ kpc and its typical width is
$\sim20$ kpc. 
Assuming a $3 \times t_{KH}$ case for a wavelength comparable to
the filament's width, it takes about $2 \times 10^9$ yrs for
the instability to act. 
This is longer than the time over which the filament is destroyed
by shocks and other instabilities and the time
needed for gas to flow through the instability region visible in
Figure~\ref{fig:RT}.
It is also possible that the filament length is relevant
which increases this timescale by factor of few. 
We therefore conclude that this structure can indeed penetrate into a
hot halo to create such a configuration.
Only much smaller modes can affect the infalling cold gas in a
filament, perhaps enhancing the surface shearing layer.

Next, we estimate the timescale for  RT instabilities \citep{murray93}
to act:
\begin{equation}\label{eq:RT}
t_{RT}=2 \pi {\left[{2 \pi g \over \lambda} \left({\rho_c-\rho_h \over
      \rho_h+\rho_c}\right) \right]}^{-1/2} \, ,
\end{equation}
where $g$ is the gravitational acceleration. Taking the acceleration at
50 kpc in MWH and a density contrast $\sim5$, we estimate the timescale for
the instability at $\sim5$ kpc scales, which corresponds to the example in
Figure~\ref{fig:RT}:  $t_{RT}\sim3.5\times 10^8$ yr, 
comparable to $t_{cool}\sim3-4\times 10^8$ yr of the photo-ionized gas in a 
warm filament at $T\sim 1-1.2\times 10^5K$ and $n_H=2 \times 10^{-4} cm^{-3}$.
Therefore, RT and cooling instabilities are acting together to form
pressure supported clouds with $T_c\sim10^4K$.    
Both of these timescales are shorter than the time over which the overdense
structure is present in the inner halo. 
This process often simultaneously forms 
many clouds in the same region, yielding
cloud complexes.  
At this stage, KH instabilities of similar wavelength could play a role. 
When clouds start to form, shearing velocities are slowed down by
pressure gradients and shocks before they accelerate to terminal
velocity so the role played by KH instabilities is
unclear.  

\section{Discussion}

Halos formed in fully cosmological environments are complex and
dynamic systems. In addition to the processes described above,
infalling satellites can be stripped of gas as they pass through
overdense filaments in the outskirts of halos, providing material for
further instabilities and cloud formation.  This effect can alleviate
the need for a high density hot medium to produce large gaseous tails
such as the one observed in the Magellanic Cloud system
\citep{putman03b}. An example of this is visible in the upper left
corners of the bottom two panels in Figure~\ref{fig:RT}.  Infalling
substructures create density waves and shocks in the halo, providing a
source of further compression and acceleration which enables RT
instabilities to develop even without the gravitational pull of the
parent halo.  Similarly, we expect density inversions and RT
instabilities to be more common in simulations with galactic outflows
(both stellar and AGN), where winds can temporarily push, compress and
accelerate both galactic and infalling material \citep{dimatteo05}.

The MWH has a growth history typical for its mass indicating that this
cloud formation process is general.  In future work we will discuss
cloud formation in larger halos where clouds form at earlier times
\citep[e.g., see Figure 6 in][]{keres09a}.  Halo clouds are likely
part of halo absorption systems observed at intermediate redshift and
some are likely analogs of the HVCs.  However, from
Figure~\ref{fig:clouds} it is already clear that many of the cold
clouds have velocities comparable to the galactic rotation which will
make searches for such objects difficult \citep{peek08}.  Cloud infall
at large radii will also likely trigger star formation and provide
kinematic enhancements in the outer disks.

Cloud formation from cooling instabilities in an idealized setting was 
simulated in \citet{kaufmann09}, demonstrating that large
entropy cores are needed for this mechanism to operate. This
is supported by \citet{binney09} who examined relevant timescales for
the onset of cooling instabilities that begin with small density
enhancements and concluded that unless the entropy profile of the halo
gas is shallow, buoyancy and conduction will prevent cloud formation. 
In our simulation clouds are seeded at intermediate radii starting from
already moderate overdensities, for which this linear analysis does
not apply.
However, the inner gas entropy and density profiles of MWH are quite
shallow, so it is not clear if slower cloud formation is prevented.
Interestingly, clouds just below our resolution limit, $< 32$
particles, do have a larger contribution from a hot gas component.  
More work is needed to determine if this is
a physical effect or purely numerical if unresolved objects partially
mix in hotter gas. In reality, even a moderate metallicity of halo
 gas could increase cloud formation efficiency. 

Both SPH and Eulerian codes have shortcomings in following the
instabilities relevant when clouds move in a hot medium. New
techniques, such as the moving mesh code of \citet{springel09}, may
avoid these shortcomings, providing new insights into the formation
and evolution of HVCs.

\subsection{Survivability of clouds}

It is interesting to determine if clouds formed in MWH can survive
their trip through the inner halo and rain onto the galaxy, providing
fresh fuel for star formation.  In the absence of magnetic fields,
thermal conduction will directly evaporate clouds smaller than about
$M_c\sim10^5 \msun$ in a free fall time for the typical radii where
clouds form in MWH \citep{mo96}.  The evolution of more massive clouds is
likely governed by a combination of RT and KH instabilities
\citep[e.g. see][]{murray93, murray04}. Our simulations do not have
sufficient resolution to follow instabilities on such small scales and
SPH itself might not be suitable for this task.

We can, however, estimate relevant timescales. Idealized simulations
show that clouds fragment when they sweep up a mass comparable to
their own \citep{murray04}.  Therefore, we integrate the descent of
clouds on radial orbits, assuming they form at 70 kpc and need to
reach 5 kpc to fall onto a galaxy.  We use $T_h=10^6K$, $T_c=10^4K$
and a density profile $\rho \propto r^{-\beta}$. We take
$\beta=0.65-0.9$ (bracketing the inner density profile of the
MWH; for simplicity we assume $T_h=$ const.), and normalize the
density at 70 kpc. 

We find that clouds above $2-3 \times 10^5 \msun$ survive if clouds
are compressed during their descent. This mass is $\sim5$ times
higher if the cloud radius remains fixed during infall.  In both cases
this is smaller than masses of the clouds formed in our
simulation.  
We also directly integrate cloud orbits accounting for
  gravity and ram pressure, assuming that clouds have angular
  momentum similar to the disk gas. This only mildly increases
  the mass limit for cloud survival compared to radial infall.  
Therefore, massive clouds seeded in the inner halo are
likely able to bring cold material to a central disk. In the non-wind
simulations at z=0 this yields $0.6 \msun$/yr of fresh gas accretion,
but if galactic outflows are included this increases to $ \sim 1
\msun$/yr (because more material is saved for later times). For the
MW, whose mass is factor of $\sim2$ larger than the simulated MWH,
the rates are likely higher, providing enough fuel for continuous star
formation over a several Gyr period.

\section{Conclusions}

Using a state-of-the-art numerical simulation of structure formation
in a \lcdm universe we follow the formation of small $\sim10^6 \msun$
cold, pressure confined clouds in a halo similar to the MW. Most
clouds are not randomly formed from small fluctuations in a hot, $T
\sim10^6$, medium as has often been supposed. Instead, they originate
from overdense, $T\sim10^5$, filamentary gas within a hot halo at
low redshift.
Such gas creates a density and entropy inversion in a gravitational field,
driving simultaneous cooling and RT instabilities.  During their
initial evolution, many clouds form from cometary structures while
later evolution cannot be followed properly with our computational
scheme. However, simple estimates indicate that a large fraction of
such clouds could indeed survive their trip to the halo center and
provide a supply of gas for late time star formation in galaxies.

We are grateful to V. Springel for allowing us to use his code and to
G. Besla, Y. Birnboim, A. Burkert and T. Cox for useful comments.

\end{document}